# A gradient model for the spatial patterns of cities

Jie Chang[1†], Guofu Yang[1†], Shun Liu[1], Hanhui Jin[2], Zhaoping Wu[1], Ronghua Xu[1], Yong Min[3], Kaiwen Zheng[1], Bin Xu[4], Weidong Luo[5], Ying Ge[1*], Feng Mao[6*], Kang Hao Cheong[7,8*]

*1 College of Life Sciences, Zhejiang University, Hangzhou 310058, China*
*2 School of Aeronautics and Astronautics, Zhejiang University, Hangzhou 310058, China*
*3 College of Computer Science and Technology, Zhejiang University of Technology, Hangzhou 310014, China*
*4 School of Public Administration, Zhejiang Gongshang University, 18 Xuezheng Street, Hangzhou, 310018, China*
*5 College of economics, Zhejiang University, Hangzhou 310058, China*
*6 School of Earth and Ocean Sciences, Cardiff University, Cardiff, CF10 3AT, United Kingdom*
*7 Science, Mathematics and Technology Cluster, Singapore University of Technology and Design (SUTD), 8 Somapah Road, S487372, Singapore*
*8 SUTD–Massachusetts Institute of Technology International Design Centre, Singapore*

*Authors for correspondence. E-mail: geying@zju.edu.cn or MaoF1@cardiff.ac.uk or kanghao_cheong@sutd.edu.sg

**To be submitted to:** *Nature*

**Abstract**

The dynamics of city's spatial structures are determined by the coupling of functional components (such as restaurants and shops) and human beings within the city. Yet, there still lacks mechanism models to quantify the spatial distribution of functional components. Here, we establish a gradient model to simulate the density curves of multiple types of components based on the equilibria of gravitational and repulsive forces along the urban-rural gradient. The forces from city center to components are determined by both the city's attributes (land rent, population and people's environmental preferences) and the components' attributes (supply capacity, product transportability and environmental impacts). The simulation for the distribution curves of 22 types of components on the urban-rural gradient are a good fit for the real-world data in cities. Based on the 4 typical types of components, the model reveals a bottom-up self-organizing mechanism that is, the patterns in city development are determined by the economic, ecological, and social attributes of both cities and components. Based on the mechanism, we predict the distribution curves of many types of components along with the development of cities. The model provides a general tool for analyzing the distribution of objects on the gradients.

**Introduction**

A fascinating event in human activities is the formation, development, expansion, and renewal of cities[1-3]. A city is composed of human beings and multiple types of functional components (or namely facilities), which are enterprises, firms, and other institutions that provide goods and services for people[4,5]. Some types of functional components, such as banks and restaurants, are concentrated near the city center. On the other hand, as a city develops manufactories tend to move outward from the city center [6,7], while some emerging industries (such as delivery stores) spring up in the urban area (Fig. 1a). The emergence, co-existence, competition, migration, and extinction of the functional components shape the spatial structure of the city[7-9]. However, the evolutionary mechanism of the city's functional spatial structure, which is composed of multiple types of components on the gradients, has not been fundamentally understood.

Each type of functional component has many individuals in a city[5,10,11]. For example, a type of fast food restaurant can be seen as a group and occupies a niche in a city, a phenomenon akin to a biological population within a community[12]. The present models for the number of components are statistical and they study the response of the local density of components—for example the number of restaurants, schools, hospitals, and banks—to the population density in a location[13-15]. These models can explain neither the whole city's structure nor the complex driving forces—apart from population—behind the distribution of components. A series of spatial economic models, pioneered by von Thünen, can explain the mechanism for the locations of land use types and functional components that affected by the land rent and transport cost gradients[3,16-18]. However, these spatial economic models have not considered the impact of people's ecological preference. A mechanical model that integrates multiple driving factors is needed urgently.

**Real-world data of the spatial pattern of components**

A city is an urban-rural system (Fig. 1a), with each component having its own distribution area. We investigate 24 types of components and choose 4 types that implement urban processes that meet the living needs of the local residents. The real-world data (Fig. 1b) shows that Kentucky Fried Chicken restaurants (KFC) are concentrated near the city center, Zhongtong express outlets (ZTO, which provides domestic parcel service) are concentrated on the outer side of KFCs, cultivated greenhouses (GH) are located in suburban areas, and dairy farms (DF) are located in ex-urban areas. Each type of component has ~$10^2$ to ~$10^3$ individuals in a city except for dairy farms, which are rare (Supplementary Table 1). The density of a type of component varied with the distance ($d$) from the city center (Fig. 1c). The univariate polynomial regressions show that, for a type of component, the density curve has a peak ($P_{max}$), which is located at a position ($d^*$) along the urban-rural gradient (Fig. 1c; Supplementary Table 2). The rank correlation showed that the $d^*$ of a type of component is related to the supply ability of the target services, which is the economic return to the investor for constructing and operating the components (Fig. 1c, d). An exception is that the $d^*$ of DF is outside GH although the net service, which is the sum of target and accompanied services, of DF is higher than GH. We find that the higher $\gamma$ value, which is the ratio of environmental impacts to target services, pushes DF outside GH (Supplementary Table 2), and we denote $\gamma$ as the 'ecological index'. The statistical model[19] we used can help us acquire the characteristics of component distribution (Supplementary Table 2), but cannot uncover the driving factors and mechanisms.

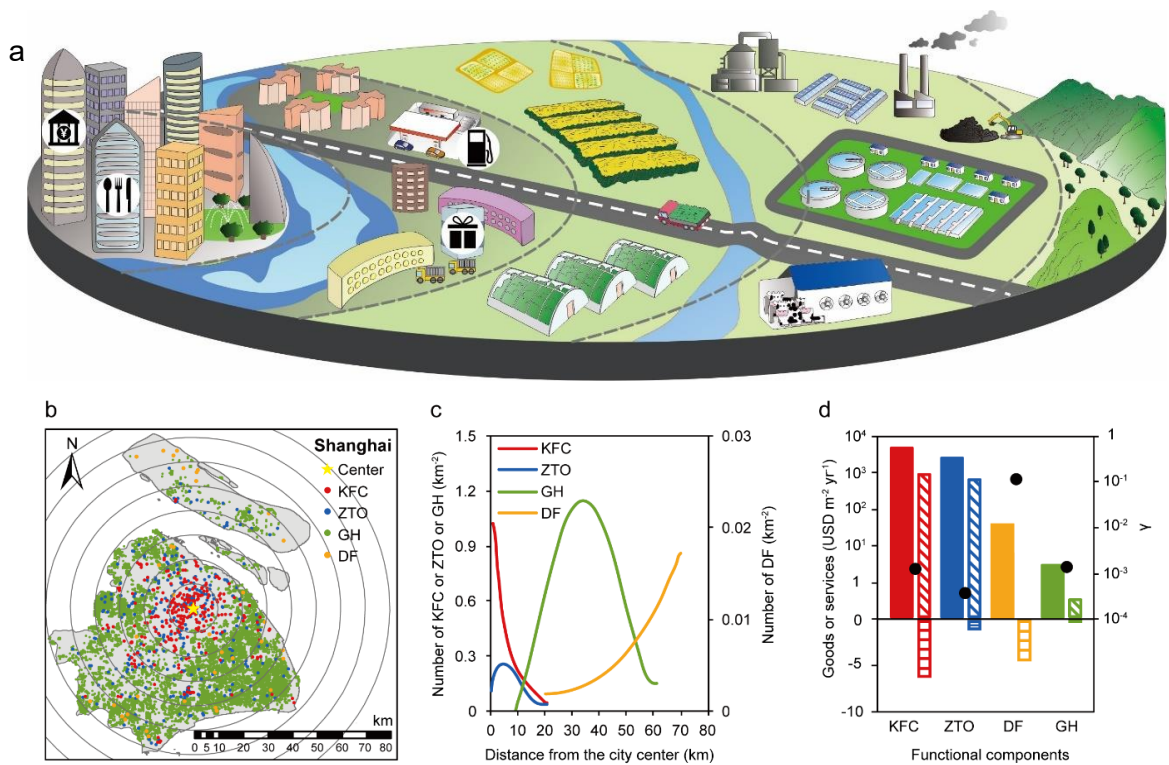

**Fig. 1 | A city's layout indicated by functional components. a,** Illustration of multiple types of components distributed from city center to rural area. **b,** Two-dimensional pattern of four types of components in Shanghai City, KFC: Kentucky Fried Chicken restaurants, ZTO: ZTO express outlets, GH: greenhouses, DF: dairy farms. **c,** One-dimensional density curves of the four types of components along the urban-rural gradient. **d,** Ecosystem services of the four types of components, solid bar: target services, diagonal bar: accompanied services, horizontal bar: dis-services, black points: $γ$ (absolute value of ratio of dis-service/target services) of the components.

**Modeling principle**

The distribution patterns of the functional components are mainly driven by economic factors[16,20]. The components, which are artificial systems, need to provide enough economic returns (target services) to the investors. The economic constraints for the distribution of functional components are the land rent and transport costs[20], while the ecological constraints are environmental impacts[6]. In an ideal city with no geographical difference[16] and only one city center, the population density monotonically decreases from the urban center to rural areas, within which the components are located and provide services for people (Fig. 2a). The two-dimensional pattern of attributes of city and components can be described as a one-dimensional density curve along the urban-rural gradient, $P(d)$, in a city (Fig. 2c, d).

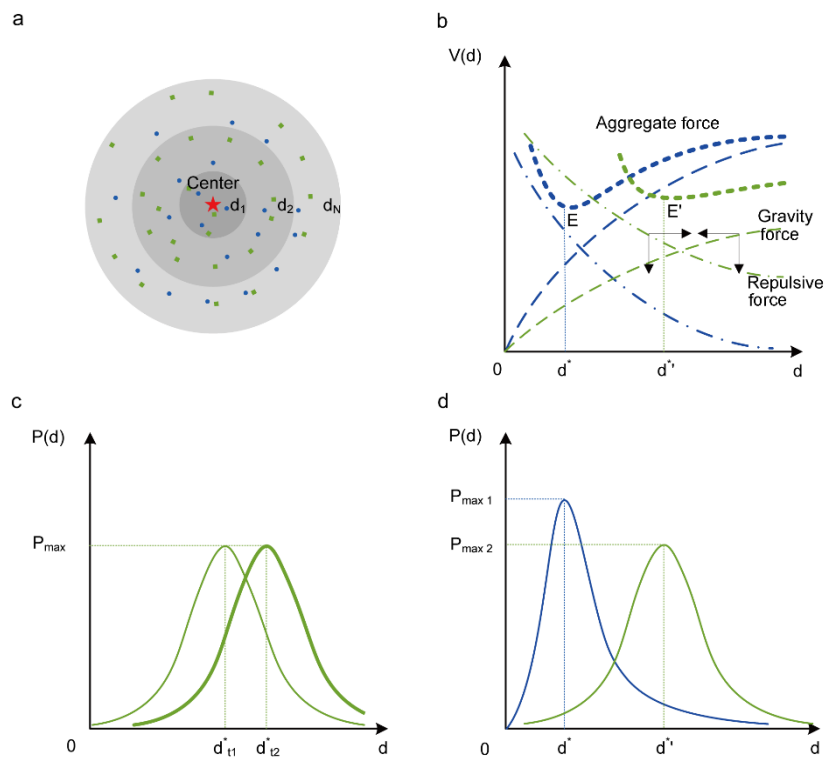

**Fig. 2 | Model hypothesis for the mechanism of spatial distribution of city components.**

**a**, In an ideal city, the two-dimensional distribution patterns of population and functional components obtain their highest density at the center and become sparser farther out. The colors of the points indicate the different types of components, and the colors of ring from dark to gray indicate the population density from high to low. **b**, Repulsive force (dash-dotted line) and reverse gravity force (thin dashed line) for components on an urban-rural gradient. Thick dashed lines show the aggregate forces. Colors of the lines corresponded the types of components in Fig 2a. $E$ and $E'$ are the minimum aggregate forces, arrows indicate the directions in which forces act on the components. **c**, Location of the density peak ($P_{max}$) of one type of component shifts along the urban-rural gradient ($d^*_{t1}$ to $d^*_{t2}$) with time. **d**, The distribution of two types of components along the urban-rural gradient. The $d^*$ and $d^{*\prime}$ of $P_{max}$ correspond to the locations of $E$ and $E'$ in Fig. 2b, colors of lines correspond to the types of components in Fig 2b.

The gravitational force ($F_g$) along an urban-rural gradient for a type of component is mainly determined by the spatial pattern of product transport cost, which is negatively correlated to the density of local population, which consumes the products provided by the functional components along the urban-rural gradient (Fig. 2b). As the transportation cost of raw materials is far less than that of the final product[20], we only consider the latter (so called "last kilometer" cost) in this study. The transportation cost depends on population density, road condition and product transportability[16,21]. Densely populated areas attract the distribution of functional components[13,22] due to a small average transportation distance ($\bar{r}$) from products to consumers (Supplementary Fig. 1a). In the city center, the population is much higher than in urban fringe and rural areas[16,21]. According to our analysis, the best bit regressions of the real-world data of population density on the distance ($d$) from city center to rural area follow the power law function (Supplementary Table 3). That is,

$$Pop(d) = Ad^\beta \qquad (1)$$

where *Pop(d)* is the average population density of the ring located at distance *d* from city center. The *Pop(d)* values were measured through random quadrat investigation on the ring. *A* is the fitted coefficient. The attenuation coefficient ($\beta$) is for the population density on the urban-rural gradient.

Based on the negative correlation between product transportation distance and population density[20], the $\bar{r}$ for a type of component at *d* ring is based on equation (1),

$$\bar{r}(d) = -\beta \ln(d) \qquad (2)$$

The transport cost also depends on the product transportability, which varied greatly among different types of components. For example, we have found that the transport of KFC's take-out products is limited to 3 km in the case cities, while the fresh vegetables in the greenhouse can be transported farther (~30 km, Supplementary Fig. 1b). We introduce Stiglitz's 'iceberg transport model' to calculate the value loss in the transportation process[23] to calculate transport cost ($C_T$),

$$C_T(d) = I\left(1 - e^{-\tau \bar{r}}\right) \qquad (3)$$

where *I* is the coefficient for the initial value of a type of product; $\tau \in (0,1)$ is the iceberg coefficient, which is the proportion of value lost per unit of distance transported; $\bar{r}$ is the average 'last kilometer' distance between the components' products and the consumers at location *d* on the urban-rural gradient.

Combining equations (2) and (3),

$$F_g(d) = I\left(1 - e^{\tau \beta \ln(d)}\right) = I\left(1 - d^{\tau \beta}\right) \qquad (4)$$

where $F_g(d)$ is the gravitational force on urban-rural gradient in a city. Equation (4) shows that the attraction is monotonic decreasing from city center to rural area.

Similar to $F_g$, the repulsive force ($F_r$) along an urban-rural gradient is monotonic decreasing from city center outward (Fig. 2b). First, high land rent (*LR*) repulses some low economic output components away from the city center[16,24]. Only the components with high economic output can be located near the city center as they can afford paying the high land rent. We have found that the land rents (*LR*) in most cities were power law decreasing on the urban-rural gradient[25],

$$LR(d) = cd^{\sigma} \qquad (5)$$

where *c* is the land rent coefficient (USD m$^{-2}$ yr$^{-1}$), and $\sigma$ is the land rent attenuation coefficient.

Second, a component's economic outputs (*m*) and people's preference for environment impacts also affect the $F_r$ for a type of component, i.e.

$$F_r(d) = LR(d) / (m/\gamma) = cd^{\sigma} / (m/\gamma) \qquad (6)$$

where *m* > 0, and $\gamma$ is the absolute value. Equation (6) means that a larger net service enables a component to be distributed near the city center, while a larger negative service pushes the component outwards (Supplementary Fig. 1c). For example, with the social prosperity in recent decades, the components with high environment impacts are moved far from the city center due to people's growing preference for better environmental quality.

The minimum of gravitational plus repulsive forces ($F_g$ + $F_r$), *E*, is changing with the development of city, and it is also different among different types of components in a city (Fig. 2b). The optimum location (*d**) corresponds to the *E* for a type of component on the urban-rural gradient where the maximum density ($P_{max}$) occurs (Fig. 2c, d). Each type of component has a *d** and some of them can overlap due to fact that they are distributed across different locations within a ring (Fig. 2a).

The total amount of a type of component and the $P_{max}$ in a city are determined by the total demand ($M$) and net service ($m$) of the component. We follow the form of Newton's gravity model and take the product of these two terms ($M \times m$) as the numerator term. Then the density of a type of component along an urban-rural gradient is

$$P(d) = G \frac{Mm}{Fg(d) + Fr(d)} \tag{7}$$

where $G$ is a coefficient to adjust the order of magnitude of the $Mm$ multiplier.

When equations (4) and (6) are substituted into equation (7), we get that the density curve of a type of component on the urban-rural gradient is

$$P(d) = G \frac{Mm}{I(1 - d^{\tau\beta}) + cd^{\sigma}/(\frac{m}{\gamma})} \tag{8}$$

The constant of proportionality ($G$) in equation (8) will be acquired via the simulation of real-world data.

**Model simulation and optimization**

We calculate the input variables using the above equations and then simulate the density curves using equation (8). The model fitting is based on the Levenberg-Marquardt algorithm, which can provide numerical solutions of nonlinear minimization (local minimum). We input the parameters and coefficients (Table 1, Extended Data Table 1, Supplementary Tables 4-5) to the nonlinear fitting module (using Origin Pro 2018, OriginLab Corporation) to estimate the coefficients ($G$, $I$) by fitting the real-world data of the 4 types of components in 13 cities. The fitting accuracy of the model parameters is evaluated and improved in each iteration, and this process continues until the accuracy of the model parameters can no longer be improved. Unfortunately, the simulated distribution density curves of KFC, ZTO and GH on the urban-rural gradient in most cities were not good fits to the data: particularly, the kurtosis did not conform to the real-world data. This means that equation (8) overlooks some factors.

**Table 1 | Input parameters and coefficients of the gradient model for four types of components in Shanghai City in different years**

| Type | Year | $M$ | $m$ | $\gamma$ | $c$ | $\sigma$ | $\tau$ | $\beta$ |
|------|------|-----|-----|----------|-----|----------|--------|---------|
| KFC | 2014 | 2.2E+09 | 2944 | 1.2E-03 | 240.1 | -0.83 | 0.3 | -0.944 |
|     | 2018 | 2.8E+09 | 2864 | 1.3E-03 | 807.5 | -0.86 | 0.3 | -0.923 |
| ZTO | 2015 | 1.2E+10 | 2045 | 5.7E-04 | 240.1 | -0.83 | 0.2 | -0.944 |
|     | 2018 | 1.5E+10 | 3236 | 5.8E-04 | 807.5 | -0.86 | 0.2 | -0.923 |
| GH  | 2010 | 2.3E+09 | 1.9 | 1.4E-03 | 240.1 | -0.83 | 0.01 | -1.074 |
|     | 2018 | 2.7E+09 | 3.3 | 1.5E-03 | 807.5 | -0.86 | 0.01 | -0.923 |
| DF  | 2009 | 2.3E+08 | 20.7 | 1.8E-01 | 240.1 | -0.83 | 0.01 | -1.074 |
|     | 2020 | 2.9E+08 | 36.3 | 1.1E-01 | 807.5 | -0.86 | 0.01 | -0.938 |

Note: The $M$-total demand of the services provided by one type of component (USD yr$^{-1}$), m- net service of one type of component (USD yr$^{-1}$ ha$^{-1}$), $\gamma$- ecological index, absolute value of ratio of disservice/target services of the component (unitless), $c$-the land rent coefficient (USD m$^{-2}$ yr$^{-1}$), $\sigma$- the land rent attenuation coefficient (unitless), $\tau$-iceberg coefficient (unitless), and $\beta$-attenuation coefficient for the population density (unitless). KFC, ZTO, GH and DF refer to Kentucky Fried Chicken shop, ZTO Express outlet, cultivated greenhouse and dairy farm, respectively. The input parameters are based on the specific city. (For details see Extended Data Table 1, Supplementary Tables 4-5.)

After re-evaluation, we introduce a new parameter, $Z$, into the denominator of the model,

$$P(d) = G\frac{Mm}{I(1-d^{\tau\beta}) + cd^{\sigma}/(\frac{m}{\gamma}) + Z} \qquad (9)$$

where the parameter *Z* represents the other factors besides the gravitational and repulsive forces in the model (Extended Data Table 2), and we dubbed it as the 'city index' due to it being related to the city attributes.

Using the modified model (equation 9), we run the fitting module again. The simulated density curves coincide with the real-world data in two periods (Fig. 3a-d; Table 1). Results showed that the fittings are significant ($R^2 > 0.72$). The density curves and the relationships between the gravitational and repulsive forces support our hypothesis in Fig. 2. Until now, we have established the gradient model. We have not further simplified the mathematical form of equation (9) because each parameter has physical significance and corresponds to the mechanism behind the distribution of components.

The *G* value of the same type of component can be obtained by the geometric mean of the same component in multiple cities,

$$G = (\prod_{i=1}^{N} G_i)^{1/N} \qquad (10)$$

where $G_i$ is the constant of proportionality in city *i*, the value of which is obtained by model fitting, and *N* is the number of the case cities.

The simulation for the density curves of the components shows that the repulsive forces to the four components are ordered KFC < ZTO < GH < DF, and so are the gravitational forces (Fig. 3e-h). The minimum points (*E*) of repulsive plus gravitational forces correspond to the *d\** (Supplementary Table 2). For example, along with the development of Shanghai City, the curve of land price and population along the urban-rural gradients tend to be gentle (absolute values of *σ* and *β* decrease), resulting in the $P_{max}$ of KFC, ZTO, GH and DF decreasing by 20%, 29%, 13%, and 21% respectively, and all the *d\** moving outwards (Fig. 3a-d). The reason is that with the development of the city, both gravitational and repulsive forces for the four components increase and the equilibrium changes (Fig. 3e-h). All the above model behaviors support our hypothesis in Fig. 2, which is that the equilibrium of forces determine the *d\**, and city's attributes determine $P_{max}$. In addition, the *d\** of GH is located at the edge of the built-up area, which reflects the shape and size of the urban area (Fig. 3c). This means that the driving force of the components quantified by the gradient model can be used to study the evolution of the scale of cities. Furthermore, the gradient model simulate many types of components co-exist in urban-rural system (Extended data Fig. 1).

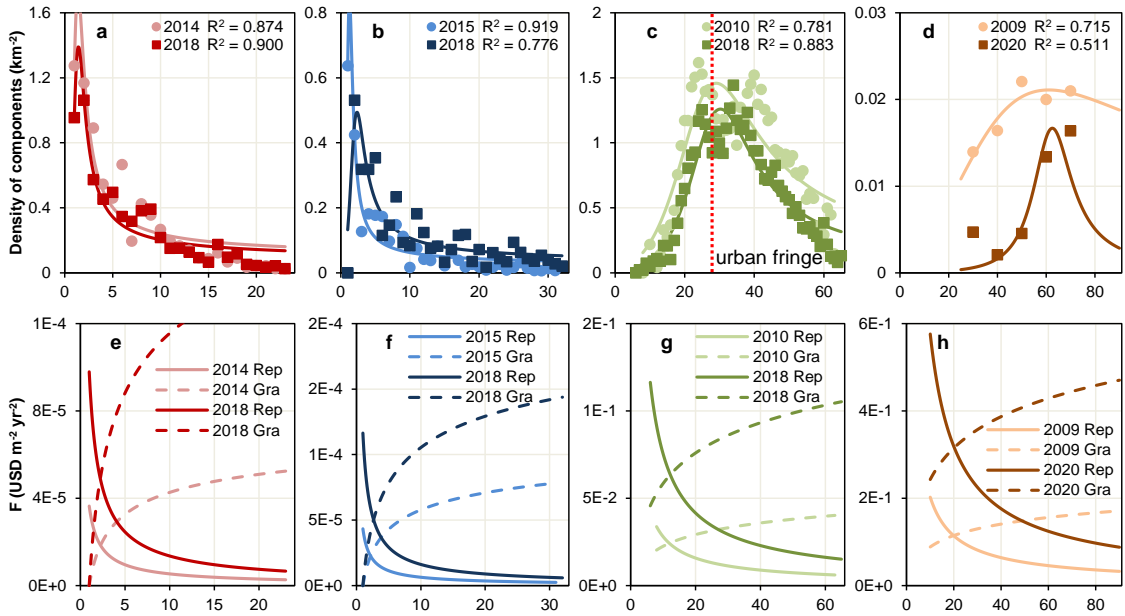

**Fig. 3 | Simulated density curves of the components and the driven forces in Shanghai City in two periods. a-d**, Simulated density curves and real-world data of the four types of components along the urban-rural gradient: **a**, Kentucky Fried Chicken shops; **b**, ZTO Express outlets; **c**, cultivated greenhouses, red dotted line indicates the location of the urban fringe; **d**, dairy farms. **e-h**, The repulsive force (solid line) and the reverse-gravitational force (dashed line) of the components

correspond to a-d, respectively.

The 4 types of components simulated by the gradient model in 13 cities showed that all the density curves of the components fitted the real-world data significantly except for KFC in Shaoxing City and ZTO in Jiaxing City, where there are insufficient individuals for simulation (Fig. 3, 4, Extended data Fig. 2). The density curves of KFC, ZTO, GH and DF were simulated well by gradient ($P < 0.05$), with average $R^2$ of 0.88, 0.81, 0.77, and 0.87 respectively. More importantly, the gradient model can simulate the density curves of multiple components (based on corresponding coefficients and parameters) in a city at the same time, and simulate their coexistence at any location within cities (Fig. 4, Extended data Fig. 3). The gradient model clearly shows the mechanism of gravitational-repulsive force shapes the specific niche of functional components on the urban-rural gradient (Supplementary Fig. 3).

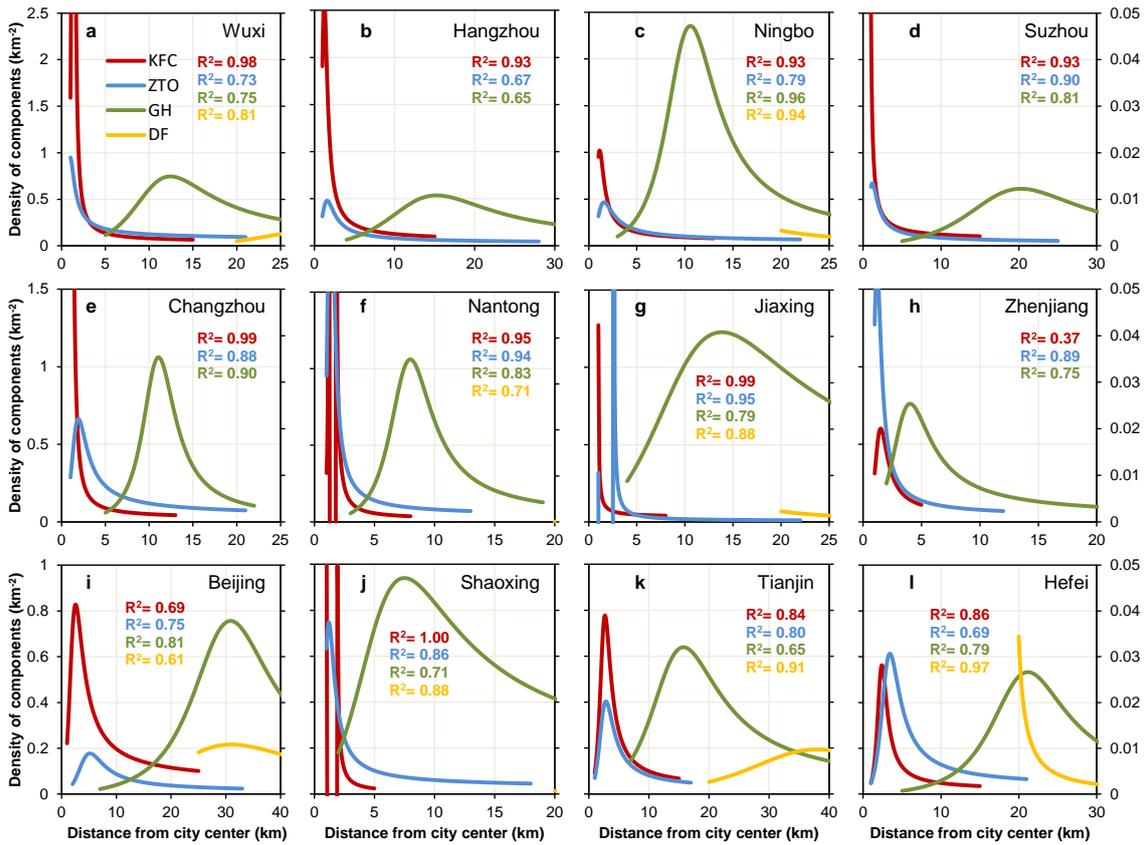

**Fig. 4 | Simulated spatial distribution of 4 types of components in 12 case cities.** The density of KFC, ZTO and GH correspond to the left axis while DF corresponds to the right axis. Note the DFs are few in some cities and non-existent in some other cities. The number of KFC in Shaoxing City and Nantong City and the number of ZTO in Jiaxing City are too small to be simulated. For the real-world data, see Extended data Fig. 3.

**Model validation**

In order to validate the robustness and the predictive power of the gradient model, we analyze the relationships between the coefficients of the model and the attributes of the cities and components. Regression analysis shows that the $Z$ of the functional components which are located in urban areas are related to urban central land price ($c$), while the $Z$ of functional components which are located in rural areas are most closely related to urban areas (Extended data Fig. 4). However, *I* is not related to any attributes. We then try to find if there are indirect relationships, and we find that the aggregate force of the model, equations (4) and (6), is derived:

$$F'(d^*) = \left[I(1 - d^{*\tau\beta}) + cd^{*\sigma}/(\frac{m}{\gamma})\right]' = 0 \quad (11)$$

where the symbols are the same as in equation (9). When the sum of gravitational and repulsive forces is the minimum, the density of components reaches the peak value, and the derivative of the sum of repulsive and gravitational forces is 0. Therefore, *I* can be calculated via

$$I = \frac{c\gamma\sigma}{\tau\beta m} d^{*\sigma - \tau\beta} \quad (12)$$

The regression analysis showed that the *d\** of a type of component was significantly related to the urban population of the 13 cities (Extended data Fig. 4). Then *I* can be derived from *d\**, which is determined by the gravitational and repulsive force. In order to study the effects of city development on component distribution, the attributes of the latest period in Hangzhou and Ningbo were used to predict the distribution of KFC, ZTO and GH over time. The predictions matched the 3 components' distribution curves well ($R^2 > 0.53$, Extended data Fig. 5).

The magnitudes of the average *G* of KFC and ZTO located in urban areas are both $10^{-12}$, while the magnitudes of the average *G* of GH and DF located outside the city are $10^{-6}$ and $10^{-8}$ respectively. The difference in the *G* values of components inside and outside the city is of 4-6 orders of magnitude (Supplementary Table 6).

Using the relationships between the model parameters and the attributes in 13 cities (Extended Data Fig. 4), we predict the density curves of the four types of components in the other two cities (Wuhan and Nanjing City). The predicted curves match well ($P < 0.05$) with the real-world data of the distribution of components (respectively, Extended data Fig. 6). The predictions validate the universality of the model among different cities (Supplementary Tables 4-5).

**Discussion**

John von Neumann used to say, "with four parameters I can fit an elephant"[26]. Our gradient model has only 3 arbitrary parameters, $G$, $I$ and $Z$, and the latter two can be calculated from real-world data (Extended data Fig. 4). It means that the gradient model is reliable and robust, for it stands on the solid ground of physics. The parameters used in gradient model are easy to observe and calculated, and the simulation results can be verified by filed investigations.

The sensitivity analysis of the gradient model shows that the distributions of the components are more sensitive to four input parameters $m$, $γ$, $σ$ and $r$ than other parameters: these four parameters largely determine the gravity and repulsion forces (Supplementary Table 7). This means that the attributes of components, including net ecosystem service ($m$) and ecological index ($γ$), as well as the attributes of cities, including land rent exponent ($σ$) and population coefficient ($β$), jointly determine the functional spatial structure of cities.

The technological innovations increase the $m$ while decreasing the $γ$ of the components. The improvements to $m$ and $γ$ raise the gravitation force and then density peak ($P_{max}$) of the components and cause the $P_{max}$ to move inward ($d*$ decreases) on urban-rural gradients. (Extended data Fig. 7). This supports the view that components have to keep up with technological innovations[10,27]. Furthermore, the response of the component density curve to $\gamma$ reveals that innovations must also reduce the environmental impact of components in order to adapt to the changes in human preference, which tends to be stricter on environment quality with increasing social prosperity[28,29]. The gradient model uncovers that the production capacity, ecological characteristics, and technological progress of a city's components have an important impact on the city's functional spatial structure.

The land rent reflects the result of competition for a scarce land resource among components[24]. The development of cities flattens the spatial patterns of land rent and population[25]. As the land rent curves become gentler, the $P_{max}$ of the four types of components decrease, but only the $d^*$ of GH and DF located outside the urban area moves outward (Supplementary Table 7). In other words, the growth of population in the peripheries of the city caused by the urbanization process (Supplementary Fig. 1b) will stimulate the $P_{max}$ of components to move outward (Extended data Fig. 7). The dynamics of cities, which include urbanization, counter urbanization, and re-urbanization, are essentially the change of the city attributes[11]. For example, the changes in economic level, population size, urban area and structure[3,7,30]. The change in total demand for the goods or services of a type of component in a city ($M$) reflects the dynamics of the population and its preferences, and it affects the $P_{max}$. The city index $Z$, which is highly related to the maximum land rent and the size of the city's urban area (Extended data Fig. 4), enhances the $P_{max}$ but maintains the position of $d^*$ (Extended data Fig. 7). Based on the above mechanism, the gradient model could help us to understand these self-organization processes of urban substructure formed from the coexistence of and competition between different types of components, and could be the basis for predicting the future city structure based on the city's attributes along with the city's development.

This model integrates the various factors and their interactions into a one-dimensional urban-rural gradient. Such a mechanism model reveals the general law of the development, expansion, and renewal of the city. The model helps to explore the limitations of the current city in structure and function, and to find out the crucial optimization points to promote sustainable development in the context of global urbanization. It can be further developed and inspire interdisciplinary studies across fields such as ecology, economics, urban planning and management, and engineering.

**Methods**

**1 Data sources**

**1.1 Selection of case cities and functional components**

The criteria for choosing case cities: (1) the data for the population and spatial pattern of population of the city is available; (2) the data for the spatial pattern of land price of the city is available; (3) a city is self-sustainable with complete processes and function. According to the criteria, we selected 15 cities in China: Shanghai, Beijing, Tianjin, Suzhou, Wuhan, Hangzhou, Nanjing, Wuxi, Ningbo, Nantong, Hefei, Changzhou, Shaoxing, Jiaxing, and Zhenjiang City, and several time periods in some cities (Supplementary Table 1).

The criteria for choosing the type of component: (1) the function of the components is to provide services to people locally, so their amount is related to the population of the city; (2) each component has a number of individuals that form a density distribution curve along the urban-rural gradient; (3) different types of components have distinguishable distribution curves along the urban-rural gradient: one type is concentrated near the city center, one type outside the center, one type is near the urban fringe, and one is in the ex-urban area. According to the criteria, we chose: (1) a type of fast food restaurant, Kentucky Fried Chicken shop (KFC), most of which are near the city center; (2) a type of express delivery outlets, ZTO delivery outlets, most of which are outside the city center, (3) a primary biological production component, cultivated greenhouses, which are mainly outside but near the suburban area, (4) a type of secondary biological production component, dairy farms, that are mainly located in ex-urban areas (Supplementary Fig. 2).

**1.2 Data of attributes of cities**

Gross domestic product (GDP) are taken from "China City Statistical Yearbook"[31] issued by the National Bureau of Statistics of China. Urban built-up areas are taken from "China Urban-Rural Construction Statistical Yearbook"[32] issued by the National Ministry of Housing and Urban-Rural Development. A city's demand for specific goods and services ($M$ in model): part of the data comes from "China Urban-Rural Construction Statistical Yearbook"[32] issued by the National Ministry of Housing and Urban-Rural Development, while the rest comes from Statistical Yearbooks of provinces and cities issued by their Statistics Bureau. These data are used to perform subsequent calculations on specific types of components.

Spatial patterns of population in a city are obtained through the WorldPop project (https://www.worldpop.org/) followed Tatem[33]. The spatial dataset is available to download in Geotiff format at a resolution of 3 arcs (approximately 100m at the equator), detail see Gaughan et al. [34]. We imported the downloaded population density raster data into a Geographic Information Systems (GIS) analysis tool (ArcGIS 10.3) for subsequent analysis. We used the regional statistical tools in ArcGIS 10.3 to process the population raster data. We set the city center as the circle center and 1 km as the radius increment to make a sequence of concentric rings that covered the entire city. Population density of each ring was calculated by using the spatial analysis toolbox in ArcGIS. The relationship between the population density of the rings of each level and the distance from the city center to the rings was obtained, and a one-dimensional distribution scatter diagram was made. The relationship between population density and distance from the city center was fitted via power function regression to make a one-dimensional distribution curve.

The spatial data of land prices in the cities in 2018, 2020 were obtained through the China Land Price Information Service Platform (http://www.landvalue.com.cn/)[35]. In October 2018, we used the real-time online function of China's land price information map service to record all standard parcel information in each case city, including land prices and standard parcel longitude and latitude, and manually reviewed the results. After that, the latitude and longitude coordinates of the standard parcels were imported into ArcGIS 10.3 (Esri corporation) for subsequent analysis. The land prices before 2018 were investigated by spatial sampling methods based on the data from the China Land Price

Information Service Platform (http://www.landvalue.com.cn/)[35]. According to the city size, we launch 4 - 8 sample lines from the city center in different directions, covering the entire city area. We record the benchmark land price of all sample points at intervals of 1 km on each sample line. The land price of these sample points is regarded as standard parcels and the distance from them to the city center are imported into excel for subsequent analysis. The city center is the commercial center, which is the place with the highest land price in the city. Land rents are calculated by dividing commercial land price by duration of land use right transfer[25]. In China, the duration is 40 years. The relationship between the land price of the rings of each level and the distance from the city center to the rings was obtained, and a one-dimensional distribution scatter diagram was made. The relationship between land rent and distance from the city center was fitted via power function regression to make a one-dimensional distribution curve.

**1.3 Data of the functional components**

The data of the revenue of KFC (USD $yr^{-1}$) for all stores in a year in China comes from Yum annual report (https://ir.yumchina.com/annual-reports). The environmental impact data comes from EIA Report (http://ishare.iask.sina.com.cn/f/byjKLlpqdW9.html). The data of the revenue of ZTO (USD $yr^{-1}$) for all stores in a year in China comes from ZTO Express annual report (https://www.zto.com/investorRelations/periodicReport.html). The environmental protection tax of food residue per ton is 3.78 USD $t^{-1}$ (the data comes from Environmental Protection Tax Law). The ecosystem services of greenhouses come from Chang et al[36]. The ecosystem services of dairy farms come from Fan et al. [37].

We obtained the spatial distribution data of these components through remote sensing images, electronic maps, and ground survey verification. In order to quantify the density of components along the urban-rural gradient, the width of concentric rings was set at 1 km for KFC, ZTO, and GH but 10 km for DF, considering the smaller number the components. The datasets for the spatial distribution of KFC, ZTO Express, and dairy farms were fetched via application programming interfaces (APIs) provided by AutoNavi Map Services (https://www.amap.com/)[38], which is the most widely used search engine and map service provider in China. We used keywords to search for points of interest (POIs). In order to ensure the accuracy of the query, in October 2018 we searched for each type of component within the administrative divisions of 15 case cities using multiple keywords with similar meanings. For example, the search for KFC used keywords such as " Kentucky Fried Chicken" or "KFC". We summarized the results of multiple keyword searches, removed obvious wrong points or duplicates, and manually reviewed the results. Then, the latitude and longitude coordinates of the four types of components were imported into ArcGIS 10.3 for subsequent data analysis.

The quadrat method in community ecology is used to determine the spatial distribution of greenhouses. We used ArcGIS 10.3 to generate a quadrat covering the entire city (the size of a quadrat is $0.5 \times 0.5$ km) to convert to KML format and import into Google Earth for visual recognition. We selected the image in 2010 as the data source. If a greenhouse is found in a quadrat, we used the landmark tool to mark in the center of the quadrat. The characteristics of greenhouses in Google Earth are: (a) bright white or bright blue, (b) regular rectangle, densely arranged with a sense of hierarchy, (c) separated by different land use types. All the quadrats containing the greenhouses were saved in KML format and imported into ArcGIS 10.3 for subsequent analysis. The number of greenhouses in each ring was recorded and divided by the area to obtain the density. Finally, the relationship between the density of greenhouses in the rings of each level and the distance from the city center to the rings was obtained, and a one-dimensional distribution scatter diagram was made.

**2. Assessment of the ecosystem services of the functional components**

**2.1 Frameworks for ecosystem services assessments for each type of component**

According to the Millennium Ecosystem Assessment (MA), ecosystem services include provisioning services, regulating services, cultural services, and supporting services[39]. In this study, the ecosystem services (goods and services) provided by artificial ecosystems (components) are divided into target services and accompanied services separately (Supplementary Fig. 4).

The target services of a type of component are determined by the investment goal of the artificial ecosystem. This means that they can be the provisioning, regulating, or cultural services in MA (2005). The target service of KFC, greenhouses, and dairy farms is the provisioning of food, which is equivalent to some of the provisioning services of natural ecosystems, while the target service of ZTO is the regulating service of distributing goods to people.

The accompanied services are equivalent to the externalities (positive or negative) in economics. They can be categorized into provisioning, regulating or cultural services. Regulating services are further divided into positive and negative (dis-services) in this paper, following the guidelines in Liu et al.[40].

The net service (NES, $m$ in model) is the sum of the ecosystem services (target service + positive regulating services + cultural services) and dis-services (environmental impacts),

$$NES = \sum_{i=1}^{n} ES_i \tag{1}$$

where $ES_i$ (USD m$^{-2}$ yr$^{-1}$) is the value of ecosystem service $i$, and $n$ is the number of ecosystem services considered in this study.

**Ecological index** ($\gamma$ in model) is calculated by the ratio of dis-services (EDS) to target goods and services (TGS) of a type of component,

$$\gamma = |EDS| / TGS \tag{2}$$

**2.2 Calculation of the ecosystem services of the components**

### 2.2.1 KFC

**Target service** A KFC store provides food as the target service, which is measured by the total value, which is calculated by dividing the company's total revenue (all stores) in the year by the total area of company stores in the country,

$$ES_p = \frac{RE_t}{N_t A} \tag{3}$$

where $ES_p$ is the provisioning services of a KFC store (USD m$^{-2}$ yr$^{-1}$), $RE_t$ is the total operating revenue of all stores of the company in one year (USD yr$^{-1}$), $N_t$ is the total number of stores belonging to the company, and $A$ is the average area of a KFC store (m$^2$).

**Accompanied services** In addition to enjoying foods, consumers in KFC stores will engage in spiritual entertainment activities such as surfing the internet, chatting, reading books, and playing games. Thus the accompanied cultural services are positive. We use the market substitution method[41] to calculate a store's cultural service. For KFC, the accompanied regulating services are mainly the environmental impacts (dis-services), including exhaust emissions, wastewater discharge, solid waste generation, and noise generation.

### 2.2.2 ZTO

**Target service** A ZTO outlet's target service is to distribute goods, which is measured by the total value, which is calculated by dividing the company's total revenue (all outlets) in the year divided by the total area of company outlets in the country,

$$ES_{r-g} = \frac{RE_t}{N_t \cdot A} \tag{4}$$

where $ES_{r-g}$ is the value of the regulating services of a ZTO outlet (USD m$^{-2}$ yr$^{-1}$), $RE_t$ is the total operating revenue of all outlets of the company in one year (USD yr$^{-1}$), $N_t$ is the total number of outlets belonging to the company, and $A$ is the average area of a ZTO outlet (m$^2$).

**Accompanied ecosystem services** Consumers will be surprised and satisfied when they get a package and open it in a delivery outlet. We use the surrogate market approach following Curtis[41] to calculate its cultural value. The accompanied regulating services are mainly the environmental impacts (dis-services), including exhaust emissions, solid waste generation, and noise generation.

### 2.2.3 Greenhouses

The investigation of ecosystem services was based on an individual greenhouse, while the intensities of services are quantified per unit area.

**Target services** The data were collected from remote sensing and field surveys, including the area of an individual greenhouse and the crop species, yield, market price of vegetables and the costs of production. A greenhouse usually grows many species of crops in one year, so the provisioning services are

$$ES_p = \frac{\sum_{sp=1}^{n} Y_{sp} \times PR_{sp}}{A} \tag{5}$$

where $ES_p$ (USD ha$^{-1}$ yr$^{-1}$) is the value of the provisioning services of a greenhouse, $Y_{sp}$ (t yr$^{-1}$) is the economic production of the crop species $sp$, $PR_{sp}$ (USD t$^{-1}$) is the average price of the crop species $sp$, and $A$ (ha) is the area of a greenhouse.

**Accompanied cultural services** The cultural services of greenhouses are positive[37]. They include aesthetics and recreation, which were not be measured in this paper.

**Accompanied regulating services** include carbon sequestration[42], soil conservation, soil fertility protection, and water saving; dis-services including soil salinization, plastic wastes, and $CH_4$ and $N_2O$ emissions[37].

### 2.2.4 dairy farms

**Target services** is the milk supply per unit farm area. The economic benefits of organic fertilizer and beef supply are ignored. The milk yield per unit farm area ($Y_m$, t ha$^{-1}$ yr$^{-1}$) is

$$Y_m = Y_{lc} \times N_{lc} / A \tag{6}$$

where $Y_{lc}$ (t head$^{-1}$ yr$^{-1}$) is the average annual yield of lactating cows, $N_{lc}$ is the number of lactating cows in a farm, and $A$ (ha) is the area of dairy farm. The value of the provisioning service ($ES_p$, USD ha$^{-1}$ yr$^{-1}$) is

$$ES_p = Y_m \times PR_m \qquad (7)$$

where $Y_m$ (t ha$^{-1}$ yr$^{-1}$) is the milk production per unit farm area, and $PR_m$ (USD t$^{-1}$) is the price of milk.

**Accompanied service** is the value of environmental impact caused by NH$_3$ and GHG emissions per unit farm area. The midpoint impacts of pollutants are calculated first, and then converted into the endpoint impacts, which is the health damage to human and ecosystems[43]. This study focuses on the impact of proximity to the cowshed, so it only focuses on the impact of gas pollutants, but the waste water and solid waste generated in the later waste treatment process are not included.

**3. Assessment of the total demand for the goods and services provided by a type of component in a city ($M$ in model).**

A city's demand for goods and services provided by the functional components is the monetary value ($M$), which is calculated according to the specific type of component. The $M$ for KFC is obtained by multiplying the city's per capita food consumption outside the home by the ratio of fast food service revenue to all catering service revenue and multiplying by the population,

$$M = CE_{f\text{-}pc} \times RA_{ff} \times POP \qquad (8)$$

where $CE_{f\text{-}pc}$ (USD person$^{-1}$ yr$^{-1}$) is the city's per capita food consumption expenditures outside, $RA_{ff}$ (%) is the ratio of fast food service revenue to all catering service revenue, and $POP$ (person) is the resident population of the city.

The $M$ for ZTO is obtained by multiplying the city's total express delivery industry revenue by the market share of ZTO and multiplying by the population,

$$M = CE_{r\text{-}pc} \times RA_{ZTO} \times POP \tag{9}$$

where $CE_{r\text{-}pc}$ (USD person$^{-1}$ yr$^{-1}$) is the city's total express delivery industry revenue, $RA_{ZTO}$ (%) is the market share of ZTO, and $POP$ (person) is the resident population of the city.

A city's demand for services provided by greenhouses is obtained by multiplying the city's per capita vegetable consumption expenditures by the population,

$$M = CQ_{v\text{-}pc} \times PR_v \times POP \tag{10}$$

where $CQ_{v\text{-}pc}$ (USD person$^{-1}$ yr$^{-1}$) is the city's per capita vegetable consumption quantity, $PR_v$ (USD kg$^{-1}$) is the price of vegetable, and $POP$ (person) is the resident population of the city.

A city's demand for services provided by dairy farms is obtained by multiplying the city's per capita fresh milk consumption expenditures by the population,

$$M = CQ_{m\text{-}pc} \times PR_m \times POP \tag{11}$$

where $CQ_{m\text{-}pc}$ (kg person$^{-1}$ yr$^{-1}$) is the city's per capita fresh milk consumption quantity, $PR_m$ (USD kg$^{-1}$) is the price of milk, and $POP$ (person) is the resident population of the city.

**4 Statistics**

The statistic functions for the attributes of cities and components with distance from the city center uses linear and nonlinear regressions (Excel 2019, Microsoft Cooperation). Linear and nonlinear regression were used to study the relationship between Z and the city attributes, and the best adjusted $R^2$ was used to select the regression form.

**5 Simulation process**

The nonlinear fitting module in Origin 2018 Pro (OriginLab Corporation) was used to simulate the spatial distribution of functional components. As the initial values of nonlinear fitting parameters may affect the simulation results, we carefully cycle through the appropriate initial values. At the beginning of the model iterative process, we use the default value of 1 as the initial value for $G$ and $I$. Meanwhile, we fix $Z$ as 0, because $Z$ only affects the $P_{max}$ without the peak position. After the model iteration is over at this stage, we relax the condition on $Z$ and continue with the iterative process to obtain the final fitting parameters.

**Acknowledgements** This work was financially supported by the National Natural Science Foundation of China (Grant No. 31870307, 31770434, 31670329). We thank Qu Z., Pan K., Zhu K, Shi M. Wong Y., Guo K., Zhang T. for the contributions.

**Author contributions** J.C. and Y.G. designed the study; G.Y., S.L., F.M., Y.M. and K.Z. developed the model; H.J., Z.W. and R.X. performed data collection, data analysis and filed investigation; B.X., W.L. undertook the economic analysis, J.C., Y.G., G.Y. and K.H.C. wrote the manuscript.

**Competing interests** The authors declare no competing interests.